\documentclass[a4paper,11pt]{article}

\usepackage{jheppub} 

\usepackage{amsmath}
\usepackage{graphicx}
\usepackage{subfigure}
\usepackage{epstopdf}
\usepackage{epsfig}
\usepackage{amssymb}
\usepackage{bm}
\usepackage{bbm}

\newcommand{\beq}{\begin{equation}}
\newcommand{\eeq}{\end{equation}}
\newcommand{\ov}{\overline}

\title{
$\bm{B\to D^{(\ast)} \tau \nu}$ and $\bm{B\to \tau \nu}$ 
in chiral $\bm{U(1)^{'}}$ models \\ with flavored multi Higgs doublets}

\author[a]{P. Ko}

\author[b]{Yuji Omura}

\author[a,c]{Chaehyun Yu}

\affiliation[a]{School of Physics, KIAS, Seoul 130-722, Korea}
\affiliation[b]{Physik Department T30, Technische Universit\"{a}t M\"{u}nchen, \\
James-Franck-Stra$\beta$e, 85748 Garching, Germany}
\affiliation[c]{SLAC National Accelerator Laboratory,  \\
2575 Sand Hill Rd, Menlo Park, CA 94025, USA}

\emailAdd{pko@kias.re.kr}
\emailAdd{yuji.omura@tum.de}
\emailAdd{chyu@kias.re.kr}

\abstract{
We discuss semileptonic and leptonic $B$ decays, $B \to D^{(*)} \tau \nu$ and 
$B \to \tau \nu$, in the chiral $U(1)'$ models which were proposed by 
the present authors in the context of the top forward-backward asymmetry 
($A^t_{\rm FB}$) observed at the Tevatron. In these models, extra Higgs doublets with nonzero 
$U(1)'$ charges are required in order to make the realistic mass matrix for up-type quarks. 
Then the extra (pseudo)scalars contribute to $A^t_{\rm FB}$ with  large flavor-changing 
Yukawa couplings involving top quark.  The contribution of the charged Higgs to $A^t_{\rm FB}$ 
is negligible,  but it may significantly affect $B$ decays: especially, 
$B\to D^{(*)}\tau \nu$ and $B \to \tau \nu$.   We investigate constraints on the $B$ decays, 
based on the recent results in BaBar and Belle experiments, 
and discuss the possibility that the allowed parameter 
region in the $B$ decays can achieve large $A^t_{\rm FB}$.}

\begin{document}

\maketitle
\flushbottom

\section{Introduction}
Recently, the BaBar collaboration analyzed semileptonic $B$ decays,
$B \to D \tau \nu$ and $B \to D^* \tau \nu$~\cite{Lees:2012xj},
and investigated the ratios of the branching ratios
for $B\to D^{(\ast)}\tau \nu$ to those for $B\to D^{(\ast)}l \nu$ ($l=e,\mu$),
\begin{equation}
R(D^{(*)}) = \textrm{BR}(B \to D^{(*)} \tau \nu)/\textrm{BR}(B \to D^{(*)} l \nu).
\end{equation}
The results are $R(D)= 0.440 \pm 0.072$ and $R(D^*)=0.332 \pm 0.030$ 
which deviate from the Standard Model (SM) predictions, $R(D)_\textrm{SM}=0.297 \pm 0.017$ \cite{Kamenik:2008tj} and $R(D^*)_\textrm{SM}=0.252 \pm 0.003$ \cite{Fajfer:2012vx} 
by $2.2\sigma$ and $2.7\sigma$, respectively~\cite{Crivellin:2012ye}.
The combined discrepancy is about $3.4 \sigma$~\cite{Lees:2012xj}, which   
might be an evidence of new physics as discussed in some recent works  
~\cite{Fajfer:2012jt, Crivellin:2012ye,Datta:2012qk,Becirevic:2012jf,Celis:2012dk, 
He:2012zp,Tanaka}.  One good candidate for new physics for this anomaly is a charged Higgs 
boson in the extended SM with extra Higgs doublets~\cite{Fajfer:2012jt, Crivellin:2012ye}.

On the other hand, a leptonic $B$ decay $B \to \tau \nu$ was measured at 
BaBar~\cite{Aubert:2008zzb} and Belle~\cite{Hara:2010dk}. The average of
the branching ratios is 
$\textrm{BR}(B \to \tau \nu)=(1.67 \pm 0.3) \times 10^{-4}$~\cite{Asner:2010qj}.  
In the SM calculation, there are some uncertainties from $|V_{ub}|$ and the $B$ meson decay 
constant $f_B$.  Still the measured number  is slightly inconsistent with the SM prediction, 
for example,  given by UTfit Collaboration, $\textrm{BR}(B \to \tau \nu)_\textrm{SM}= 
( 0.84 \pm 0.11 ) \times 10^{-4}$ ~\cite{Bona:2009cj}. 
The Belle experiment recently presented a new result,
$\textrm{BR}(B\to \tau\nu)= 0.72^{+0.27}_{-0.25}\pm 0.11$ by making use of a hadronic
tagging method for $\tau$ decays with the full data sample~\cite{Adachi:2012mm},
and the combined average is consistent with the SM prediction.
Since both semileptonic and leptonic $B$ decays may be affected by the same 
new physics  (e.g., charged Higgs boson),  such new physics scenario will be 
strongly constrained  by combined analysis of the $B$ decays.

An interesting point  is that it is difficult for the so-called type-II 
two-Higgs-doublet model (2HDM), which could be well motivated by minimal 
supersymmetric SM (MSSM), to explain the discrepancies in $R(D)$ and 
$R(D^{*})$~\cite{Lees:2012xj}.  
Other types of 2HDMs with natural flavor conservation, where Yukawa 
couplings are controlled by $Z_2$~\cite{Glashow} or $U(1)_H$~\cite{Ko-2HDM} 
symmetry, also allow only the so-called minimal flavor violation (MFV) in the charged Higgs sector, and it is impossible to accommodate  $R(D)$ and $R(D^*)$ at BaBar 
simultaneously~\cite{Fajfer:2012jt, Crivellin:2012ye,He:2012zp,Tanaka}. 
Eventually, we would have to introduce non-minimal-flavor-violating (non-MFV) 
terms in order to explain the $R(D)$ and $R(D^{*})$ data, which would tend to
generate too large flavor changing neutral currents (FCNCs).   
It would be highly nontrivial to introduce non-MFV interactions in the Yukawa couplings 
in 2HDMs without too excessive flavor violations  in the $K$ and $B$ meson sectors. 

In ref.~\cite{Ko-Top,Ko-Top2}, the present authors proposed flavor models with extra 
Higgs fields, where gauged $U(1)'$ symmetry forbids the potentially problematic 
FCNCs for $B$-$\bar{B}$ and $K$-$\bar{K}$ mixings but allows certain amounts of  
FCNCs which are still consistent with experimental data, by slightly breaking 
the criteria of  ref.~\cite{Ko-2HDM}.
There, neutral CP-even scalars and a CP-odd scalar can have large $(t,q)_{q=u,c,t}$ 
elements of Yukawa couplings. Their tree-level mediations 
enhance the top forward-backward asymmetry $(A^t_{\rm FB})$ at the Tevatron,
while accommodating with the newest strong constraints from LHC
thanks to the destructive interference among the scalars~\cite{Ko-Top,Ko-Top2,Ko-LHC,Ko-LHC2}. 
In previous works, phenomenology of the charged Higgs boson in our models 
was not considered carefully because the charge Higgs boson cannot have 
sizable contribution to the top quark production at hadron colliders.
However, couplings of the charged Higgs boson to the bottom quark may be large 
so that the models could be strongly constrained by the $B$ decays. 
For example, the $(b,u)$ element of the charged Higgs coupling,
which is constrained by $B \to \tau \nu$,  may become large if the $(t,u)$ element 
of the Yukawa coupling of the pseudoscalar boson is large.
Besides, the $(b,c)$ element, which modifies the branching ratios 
of $B \to D^{(*)} \tau \nu$, could be large, if the $(t,c)$ element of the pseudoscalar 
Yukawa coupling is large (see eq.~(\ref{vijd}), for example).   
In this paper, we first  investigate if our  $U(1)^{'}$ flavor models can explain 
the discrepancies in the $B$ semileptonic decays while keeping consistency
with $\textrm{BR}(B\to \tau \nu)$. Then we discuss the possibility that
the allowed parameter regions can achieve the large enough $A^t_\textrm{FB}$ 
which was observed at the Tevatron.

This paper is organized as follows. In section~\ref{sec:Setup}, we give a brief review 
of our models proposed in ref.~\cite{Ko-Top,Ko-Top2}. We assign family dependent 
$U(1)'$ charges to the right-handed (RH) up-type quarks in order to generate 
flavor changing $Z'-u_R-t_R$ couplings in the mass eigenstates.  Then we will immediately realize that it is mandatory to introduce extra $U(1)'$-charged 
Higgs doublets in order to write Yukawa couplings for the up-type quarks including 
top quark, which have been first realized in ref.~\cite{Ko-Top,Ko-Top2}. 
Thus we are led to multi-Higgs doublet models in the presence of a new spin-1 
$Z'$ boson with chiral couplings to the SM fermions. 
Sections~\ref{sec:2HDM} and \ref{sec:3HDM} are devoted to discussion of  the phenomenology for $B \to D^{(*)} \tau \nu$ and $B \to \tau \nu$ in our 2HDM and 
three-Higgs-doublet model (3HDM), respectively.
We also give comments on the constraints on the charged Higgs boson from the 
$B \to X_s \gamma$  process in section~\ref{sec:constraints}.
Finally, we summarize our results in section~\ref{sec:summary}.
The general Higgs potential in multi Higgs doublet models is discussed in 
appendix~\ref{sec:potential}.

\section{Models with extra Higgs and gauged flavor $U(1)'$ }
\label{sec:Setup}
Adding extra Higgs doublets to the SM is one of the interesting extensions to the SM. 
However such an extension generically suffers from neutral Higgs-mediated 
FCNC problem at tree level,  if up- and down-type quark masses get contributions 
from the vacuum expectation values (VEVs) of more than one Higgs doublets. 
For example, when the mass matrix of up-type quarks ($M^u_{ij}$) depends on
two Higgs doublets, $(H_1,H_2)$,  as 
\[
(M^u)_{ij}=(y_1)_{ij} \langle H_1 \rangle+(y_2)_{ij} \langle H_2 \rangle,
\] 
all scalar components of $(H_1,H_2)$ would have flavor-dependent couplings.
This is because $(y_1)_{ij}$ and $(y_2)_{ij}$ cannot be diagonalized  
simultaneously without any relation between $(y_1)_{ij}$ and $(y_2)_{ij}$. 
A simple way to control the flavor structures of Yukawa couplings is to assign 
a symmetry to extra Higgs and matter fields.  The most popular symmetry is 
an extra $Z_2$ symmetry which is softly broken~\cite{Glashow}.    
In ref.~\cite{Ko-2HDM,Ko-Top,Ko-Top2},  gauged $U(1)'$ symmetry is 
assigned to the SM fermions and newly introduced extra Higgs doublets  
instead of $Z_2$ symmetry. 
Especially, in ref.~\cite{Ko-Top,Ko-Top2}, only right-handed (RH) up-type  quarks are 
charged flavor-dependently under $U(1)'$, and  FCNCs involving only top quark 
can be enhanced.  The authors constructed both 2HDM and 3HDM  
depending on the $U(1)'$ charge assignments to the SM fermions. 

When only the RH up-type quarks are charged in a flavor dependent way,  the extra  Higgs doublets charged under $U(1)'$ couple with the RH up-type quarks in the form
\begin{equation}
V_y= y^u_{ij} \overline{Q_i} \widetilde{H_j} U_{Rj} 
+y^d_{ij} \overline{Q_i} H_2 D_{Rj} + y^l_{ij} 
\overline{L_i}  H_2 E_{Rj}  + h.c. .
\end{equation}
Here $H_j (j=u,c,t)$ are charged under $U(1)'$, but $H_2$ is  neutral under 
$U(1)'$ and has the same quantum numbers as the SM Higgs doublet.
The SM leptons and the down-type quarks get masses from VEV of $H_2$. 
Note that the up-type quarks cannot have gauge invariant mass terms without 
new $U(1)'$-charged Higgs doublets $H_j$.

In general, there may be up to four Higgs doublets: $H_2$ and $H_{u,c,t}$. 
The actual number of Higgs doublets will depend on the $U(1)'$ charge 
assignment. Motivated by the $A^t_{\rm FB}$ at the Tevatron, 
we use the charge assignment $(u_j)=(0,0,1)$ on the RH up-type quarks
in the 2HDM, where we identify $H_u$ and $H_c$ as $H_2$ and $H_t$ as $H_1$,
respectively. In the 3HDM, we use the charge assignment $(u_j)=(-1,0,1)$
and we identify $H_u$ as $H_1$, $H_c$ as $H_2$, and $H_t$ as $H_3$, 
respectively.\footnote{There could be other assignment of $(u_j)$ but we consider 
only these two cases for simplicity.}
In addition to the extra Higgs doublets $H_j$,  a SM singlet scalar 
with nonzero $U(1)'$ charge, $\Phi$, may exist 
in order to break $U(1)'$. In general, its CP-even component mixes
with CP-even components of $H_j$
after electroweak and $U(1)'$ symmetry breaking.

In the following two sections, we will concentrate on these 2HDM and 
3HDM. Then we investigate the impact of the charged Higgs boson sector on 
the (semi)leptonic $B$ decays and discuss the results in the context 
of the $A^t_{\rm FB}$ at the Tevatron. 

\section{2HDM\label{sec:2HDM}}
In this section, we consider the chiral $U(1)'$ flavor model
with the charge assignment $(u_i)=(0,0,1)$ 
in the interaction eigenstates\footnote{We use the same notations 
as in ref.~\cite{Ko-Top,Ko-Top2}.}.
There are a neutral Higgs doublet, $H_2$, and $U(1)'$-charged Higgs doublet, $H_1$,
as we discussed in section~\ref{sec:Setup}.

\subsection{Yukawa couplings}
In the 2HDM case, there are three CP-even neutral scalars 
$(h_1,h_2,h_3)$,  one CP-odd pseudoscalar $(a)$, and one charged Higgs 
pair $(h^{\pm})$ after electroweak and $U(1)'$ symmetry breaking. 
Note that there is one more CP-even scalar from  
the $U(1)'$-charged singlet scalar $\Phi$ compared with the usual 2HDM.
 
Let us define the Yukawa couplings for the neutral scalar bosons in the mass bases 
as follows: 
\begin{equation}
V_{y^u}=Y^{u(k)}_{ij} h_k \ov{u_{Li}} u_{Rj} 
-iY^{au}_{ij} a \ov{u_{Li}} u_{Rj} + h.c..
\end{equation}
$Y^{u(k),au}_{ij}$  depend on the Higgs VEVs and 
the diagonalizing matrices for the quark mass matrices.  
The Yukawa couplings can be derived in a straightforward manner.
For example, the Yukawa couplings of the lightest CP-even scalar boson $h_1$ 
and pseudoscalar boson $a$ are given by    
\begin{eqnarray}
Y^{u(1)}_{ij}&=& \frac{m_i^u \cos \alpha}{v \cos \beta} \cos \alpha_{\Phi}
\delta_{ij}+ \frac{2  m_i^u}{v\sin 2 \beta} (g^u_R)_{ij} \sin (\alpha-\beta)
\cos \alpha_{\Phi} , 
\label{eq:Yuij_1}
\\
Y^{au}_{ij}&=&  \frac{m_i^u \tan \beta}{v } \delta_{ij}
-  \frac{2 m_i^u}{v\sin 2 \beta}  (g^u_R)_{ij} .
\label{eq:Yuij_a}
\end{eqnarray}
The parameters $v$ and $ \beta$ are defined by $( \langle H_1 \rangle,\langle H_2 \rangle)
=(v \cos \beta /\sqrt{2},v \sin \beta /\sqrt{2})$, and $\alpha$ and $\alpha_{\Phi}$
are the mixing parameters among  the $3$ CP-even neutral scalars.

In the above equations, the mixing matrix $g_R^u$ is defined as ~\cite{Ko-Top,Ko-Top2} 
\begin{equation}
(g^u_R)_{ij}=(g^u_R)^*_{ji} = (R_u)_{ik} u_k (R_u)^*_{jk}, 
\label{eq:gu_R} 
\end{equation}
where the matrix $(R_u)_{ij}$ is defined by 
\[
(M^{\dagger}_uM_u)_{ij} = (R_u)^{\dagger}_{ik} (m_u^2)_k  (R_u)_{kj} . 
\] 
For the 2HDM case with $(u_k) = (0,0,1)$,  $g_R^u$ is reduced to 
\[
(g^u_R)_{ij}=(g^u_R)^*_{ji} = 
(R_u)_{i3} (R_u)^*_{j3} .
\]
Note that the neutral Higgs bosons have flavor-dependent couplings to the up-type quarks (see the second terms in eq.~(\ref{eq:Yuij_1}) and (\ref{eq:Yuij_a})).
The flavor-dependent couplings of $Z'$ which is the gauge boson of $U(1)'$
are also linear in $(g^u_R)_{ij}$ in our 2HDM~\cite{Ko-Top,Ko-Top2}.
These flavor-dependent couplings make additional contributions 
to $A^t_{\rm FB}$ at the Tevatron,  $t\bar{t}$ cross section, and the same-sign 
top-quark pair production cross section.
Recently, the CMS collaboration has announced stringent bounds
on the cross section for the same-sign top-quark pair production: 
$\sigma^{tt} \le 0.39$ pb at 95\%
confidence level~\cite{Chatrchyan:2012sa,cmsnewtt}.
This strong bound excludes simple scenarios such as the original $Z'$ model 
that only one mediator contribution is taken into account.  However, flavor models 
usually have several mediators to couple with the SM  particles flavor-dependently, 
and they interfere with each other. In fact, the pseudoscalar,  CP-even scalars, 
and $Z'$ have destructive interference  in our $U(1)'$ flavor models and 
we could find 
the points  evading the stringent upper bound on the same-sign top-quark pair 
production while enhancing $A^t_{\rm FB}$~\cite{Ko-Top,Ko-Top2,Ko-LHC,Ko-LHC2}.

The origin of this non-MFV couplings in our models is the flavor-dependent 
$U(1)'$ interactions of $Z'$.  
(Recall that  $\left( g_R^u \right)_{ij} \propto \delta_{ij}$ if the $U(1)'$
were flavor-independent, i.e., $(u_k) \propto (1,1,1)$.)  
Thus our 2HDM is a nice realization of non-MFV models where the flavor 
non-universality has its origin in the flavor-dependent $U(1)'$ gauge interactions. 
The amount of flavor non-universality is related with the local gauge symmetry and 
its spontaneous breaking. 
It is worthwhile to emphasize that the nature of 
flavor non-universality is neither completely arbitrary nor {\it ad hoc}. 
These are very unique features of the 2HDM (and 3HDM described in the next section) 
we have proposed in ref.~\cite{Ko-Top,Ko-Top2}. 

Similarly to the neutral Higgs Yukawa couplings, the charged Higgs Yukawa 
couplings are defined by  
\begin{equation}
V_c =
 -Y^{-u}_{ij} h^- \ov{d_{Li}} u_{Rj}+ Y^{+d}_{ij} h^+ \ov{u_{Li}} d_{Rj} 
+ Y^{+l}_{ij} h^+ \ov{\nu_{Li}} e_{Rj}
+ h.c..
\end{equation}
There are definite relations between Yukawa couplings of the pseudoscalar  boson 
($a$) and those of the charged Higgs bosons ($h^\pm$):  
\begin{eqnarray}\label{eq:relation}
Y_{ij}^{-u} &=& \sum_l V^*_{li} Y^{au}_{lj} \sqrt{2}, \nonumber \\ 
Y_{ij}^{+d} &=& \sum_l V_{il} Y^{ad}_{lj} \sqrt{2},  \label{vijd} \\
Y_{ij}^{+e} &=& \sum_l (V_\textrm{PMNS})_{il} Y^{ae}_{lj} \sqrt{2},   \nonumber 
\end{eqnarray}
where $V_{ij}$ is the CKM matrix and
$(V_\textrm{PMNS})_{ij}$ is the Pontecorvo-Maki-Nakagawa-Sakata matrix.

In order to accommodate the large $A^t_{\rm FB}$ observed at the Tevatron, 
large flavor-changing Yukawa couplings of the pseudoscalar are inevitable
~\cite{Ko-Top,Ko-Top2}. As the authors pointed out in ref.~\cite{Ko-Top,Ko-Top2,Ko-LHC,Ko-LHC2}, the pseudoscalar contribution to the same-sign top-quark pair production has 
 destructive  interference  with other neutral scalar and $Z'$ contributions.
Once we consider a large FCNC in the pseudoscalar 
Yukawa couplings in the up-type quark sector in order to enhance 
the $A^t_{\rm FB}$, 
it is mandatory to have large flavor-changing couplings 
in the charged Higgs sector, which would affect various  phenomenology including 
the $B$ meson system.  

In the following, we will discuss phenomenology of the charged Higgs boson.

\subsection{ $R(D^{(*)})$ and $\textrm{BR}(B\to\tau \nu)$ in 2HDM}

The charged Higgs boson in our model will contribute to the (semi)leptonic 
$B$ decays and would modify the SM lepton universality which is a result of the 
$W^\pm$ contributions derived from the underlying $SU(2)_L$ gauge theory.  
Since the charged Higgs contributions  are proportional to the final lepton mass, 
we consider the $h^\pm$ contributions  only to $B \rightarrow D^{(*)} \tau \nu$ 
and $B \rightarrow \tau \nu$, and use the SM predictions for  other leptonic channels. 

In the effective Hamiltonian approach, the semileptonic decays 
$B\to q \tau \nu$ and leptonic decay $B\to \tau \nu$ are described 
by the effective Hamiltonian~\cite{Crivellin:2012ye},

\begin{equation}
H_\textrm{eff}= C^{qb}_\textrm{SM}(\overline{q_L} \gamma_\mu b_L)
(\overline{\tau_L}\gamma^\mu \nu_L) 
+C_R^{qb}(\ov{q_L}b_R) (\ov{\tau_R} \nu_L)
+C_L^{qb}(\ov{q_R}b_L )(\ov{\tau_R} \nu_L),
\end{equation}  
where $q=u,c$ is the up-type quark flavor.  In the above equation, 
$C^{qb}_\textrm{SM}$ is the Wilson coefficient for the $W$ exchange
in the SM and $C^{qb}_{R,L}$ are those for the charged Higgs exchange present 
in our models. $R(D^{(*)})$ and $\textrm{BR}(B\to\tau \nu)$ are given by the 
following expressions depending on the above Wilson 
coefficients~\cite{Crivellin:2012ye}: 
\begin{eqnarray}
R(D)&=& R_\textrm{SM} 
\left (1+ 1.5 ~\textrm{Re} \left ( \frac{C^{cb}_{R}+C^{cb}_{L}}{C^{cb}_{SM}} \right ) 
+ \left|  \frac{C^{cb}_{R}+C^{cb}_{L}}{C^{cb}_{SM}} \right |^2 \right ), \\
R(D^*)&=& R^*_\textrm{SM} 
\left (1+ 0.12 ~\textrm{Re} \left ( \frac{C^{cb}_{R}-C^{cb}_{L}}{C^{cb}_{SM}} \right )
+ 0.05 \left | \frac{C^{cb}_{R}-C^{cb}_{L}}{C^{cb}_{SM}} \right |^2  \right ),\\
\textrm{BR}(B \rightarrow \tau \nu ) &=& \frac{G^2_F |V_{ub}|^2}{8 \pi} m^2_{\tau} 
f_B^2 m_B \tau_B \left (1- \frac{m^2_{\tau}}{m^2_B} \right )^2 
\left |1+ \frac{m^2_B}{m_b m_{\tau}} 
\left ( \frac{C^{ub}_{R}-C^{ub}_{L}}{C^{ub}_{SM}} \right ) \right |^2,
\end{eqnarray}
where each Wilson coefficient is at the $B$ meson scale~\cite{Datta:2012qk}.
Here, $R^{(*)}_\textrm{SM}$ are given by 
$\textrm{BR}(B\to D^{(*)}\tau \nu)/\textrm{BR}(B\to D^{(*)}l\nu)$ 
in the SM.

Integrating out the heavy degrees of freedoms ($h^\pm$) in our 2HDM, 
we can obtain the Wilson coefficients at the charged Higgs mass scale: 
\begin{eqnarray}
C^{qb}_\textrm{SM}&=&2V_{qb} (V_\textrm{PMNS})^*_{\nu \tau}/v^2, \\
\label{eq:CL}%
\frac{C^{qb}_L}{C^{qb}_\textrm{SM}}&=& 
\frac{m_qm_{\tau}}{m^2_{h^{+}}} \tan^2 \beta 
-  \sum_l \frac{V_{lb}}{V_{qb}} \frac{m_l^u m_{\tau}  
(g^u_R)_{lq} }{m^2_{h^{+}} \cos^2 \beta},
\\
\label{eq:CR}%
\frac{C^{qb}_R}{C^{qb}_\textrm{SM}}&=&
- \frac{m_bm_{\tau}}{m^2_{h^{+}}} \tan^2 \beta. 
\end{eqnarray}
We note that $C^{qb}_R/C^{qb}_\textrm{SM}$ is flavor-blind, but
$C^{qb}_L/C^{qb}_\textrm{SM}$ depends on the flavor, $q=u$ or $c$. 
If $(g^u_R)_{ij}=\delta_{ij}$ were satisfied, $C^{qb}_{L,R}$ would be the same 
as the type-II 2HDM and one could not accommodate the $R(D)$ and $R(D^*)$ 
simultaneously. However in our flavor-dependent $U(1)'$ models,  we have 
$(g^u_R)_{ij} \neq \delta_{ij}$ and  there appears  a new term in eq.~(\ref{eq:CL})  
which is absent in the type-II 2HDM. This new term may give rise to a possibility 
to accommodate  $R(D)$ and $R(D^*)$ unlike the type-II 2HDMs. 
As discussed in the previous subsection,  we could generate this non-MFV interactions from flavor-dependent $U(1)'$ gauge couplings, which are very interesting aspects  
of our multi-Higgs doublet models with flavor-dependent $U(1)'$ gauge interactions. 

The contribution of the charged Higgs boson to the semileptonic $B$ decays, 
$B\to D^{(\ast)}\tau \nu$, is controlled by  $C^{cb}_{L,R}$ while that to the leptonic 
$B$ decay, $B\to \tau \nu$, is affected  by $C^{ub}_{L,R}$. 
As discussed in ref.~\cite{Lees:2012xj,Fajfer:2012jt, Crivellin:2012ye,
Datta:2012qk,Becirevic:2012jf}, it is impossible to explain $R(D^{(\ast)})$ 
and $\textrm{BR}(B\to \tau\nu)$ simultaneously within the 2HDMs with MFV, 
where the Yukawa couplings are fixed 
by the angles $\alpha$ and $\beta$ up to the quark mass.  
However, if $C^{cb}_{L,R}$ and $C^{ub}_{L,R}$ are mutually independent 
as in our models, both the measured values of $R(D^{(\ast)})$ and 
$\textrm{BR}(B\to\tau\nu)$ might be explained by taking appropriate parameters 
for the Wilson coefficients~\cite{Crivellin:2012ye}.
 
In ref.~\cite{Ko-Top,Ko-Top2}, the present authors suggested that large $Y^{au}_{tu}$ is 
required in order to achieve the large $A^t_{\rm FB}$ and evade the strong bound 
from the same-sign top-quark pair production signal.
Such large $Y^{au}_{tu}$  can be realized by large $(g^u_R)_{tu}$ 
and small $\sin 2 \beta$, but it leads to large $Y^{-u}_{bu}$  
according to eq.~(\ref{eq:relation}) in our 2HDM. Therefore the charged Higgs 
contributions to the (semi)leptonic $B$ decays could be too large, and have to
be studied carefully. 

In numerical analysis for (semi)leptonic $B$ decays in our 2HDMs, 
we take the following parameter regions:
$1 \le \tan \beta \le 100$,
$200~\textrm{GeV} \le m_{h^+} \le 1~\textrm{TeV}$,
and
$0\le |(g_R^u)_{tu}|, |(g_R^u)_{tc}| \le 1$, respectively.
We use the input parameters in the SM which are given by the global fit~\cite{CKMfitter}. 

\begin{figure}[!t]
\begin{center}
{\epsfig{figure=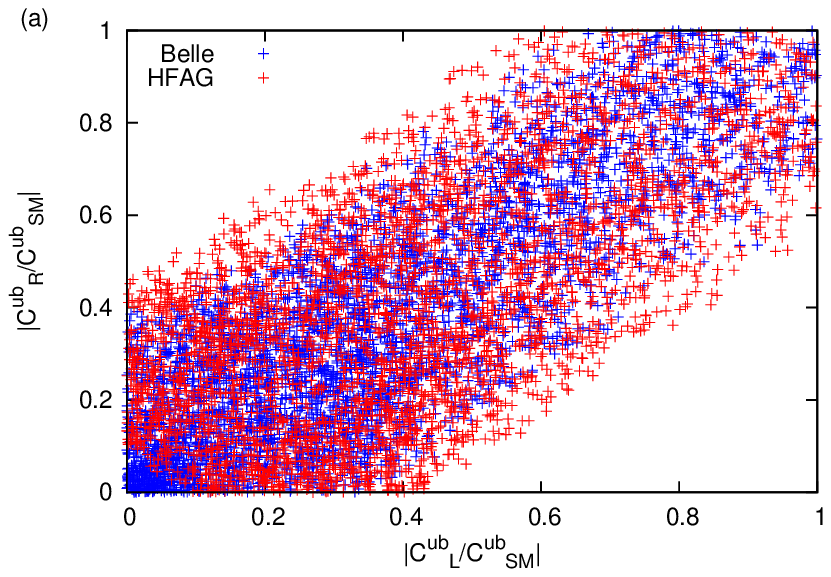,width=0.5\textwidth}}{\epsfig{figure=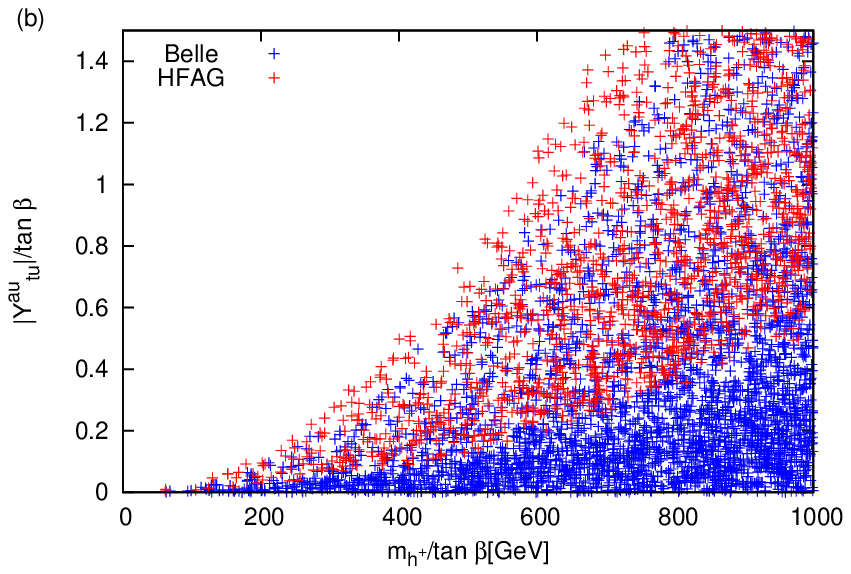,width=0.5\textwidth}} 
\end{center}
\vspace{-0.5cm}
\caption{
Bounds on Yukawa couplings and $m_{h^+}/\tan\beta$ 
from $B \to \tau \nu$. 
The red region is the allowed region for the combined data~\cite{Asner:2010qj} 
without the recent Belle data and the blue one is allowed 
by the recent Belle experimental result~\cite{Adachi:2012mm}.
We used the relation $Y^{-u}_{bu}=\sqrt{2} V^{*}_{tb} Y^{au}_{tu}$, ignoring the other elements of the Yukawa coupling.
}
\label{fig:2HDM-Btaunu}
\end{figure}

First, we consider only the $B\to \tau \nu$ decay.
In figure~\ref{fig:2HDM-Btaunu} (a), 
we show constraints on 
$|C^{ub}_L/C^{ub}_\textrm{SM}|$ and $|C^{ub}_R/C^{ub}_\textrm{SM}|$.
The red points are allowed by the combined data 
by Heavy Flavor Average Group (HFAG)~\cite{Asner:2010qj}
without the recent Belle data in the $1\sigma$ level. 
The blue points are consistent with 
the recent Belle data on $B\rightarrow \tau \nu$ in the $1\sigma$ level
~\cite{Adachi:2012mm}.
The SM point of $C^{ub}_L/C^{ub}_\textrm{SM}=C^{ub}_R/C^{ub}_\textrm{SM}=0$
is slightly deviated from the red region,  but it is in good agreement with the recent
Belle data.  If the new Belle data is combined together with the old data,
the small discrepancy in $\textrm{BR}(B\to \tau \nu)$ might disappear.
Then, $\textrm{BR}(B\to \tau\nu)$ will give a strong constraint 
on the parameters related with charged Higgs boson. 
In figure~\ref{fig:2HDM-Btaunu} (b), we depict the allowed regions in our 2HDM for 
$m_{h^+}$  and $|Y^{au}_{tu}|$, with both being scaled by $\tan \beta$.
The red and blue regions are consistent with
the combined data and new Belle data, respectively.\footnote{
By using the averaged value of the combined data by HFAG and the new Belle data
for $\textrm{BR}(B\to \tau \nu)$, one may obtain similar results. 
But there is no official averaged value up to now~\cite{Y.Kwon} and we considered
the new Belle data separately in this paper.}

In our $U(1)'$ models, there exists an additional gauge boson $Z^\prime$.  
One of the Higgs fields is charged under the extra $U(1)'$, so that
the $\rho$ parameter deviates from the SM prediction at tree level by 
\begin{equation}
\label{eq:Constraint-Rho}
 \Delta \rho_{{\rm tree}} =\{ h_i (\langle H_i \rangle \sqrt{2} /v )^2 \}^2
 \frac{ g'^2}{ g^2_Z}\frac{  m_{\Hat{Z}}^2 }{ m_{\hat{Z}'}^2 - m_{\Hat{Z}}^2},
\end{equation}
where $ m_{\Hat{Z}}^2 = g_Z^2 v^2$ and 
$ m_{\hat{Z}'}^2= g'^2  v^2 \{h_i^2 (\langle H_i \rangle \sqrt{2} /v )^2 \}
+g'^2  h^2_{\phi} v_\phi^2 $.  This form can be applied 
to 2HDM, 3HDM and etc., fixing the charge assignment, $\{h_i \}$ of $H_i$
and $h_\phi$ of $\Phi$. 
In the case of our 2HDM, $h_i (\langle H_i \rangle \sqrt{2} /v )^2= \sin^2 \beta$.
Therefore, the tree-level $\rho$ parameter favors small $\tan\beta$ region
 ~\cite{Ko-2HDM}.   This in turn implies that $Y^{au}_{tu}$ of $O(1)$ can be realized 
for $m_{h^+}/\tan \beta$ of $O(100)$ GeV.

\begin{figure}[!t]
\begin{center}
{\epsfig{figure=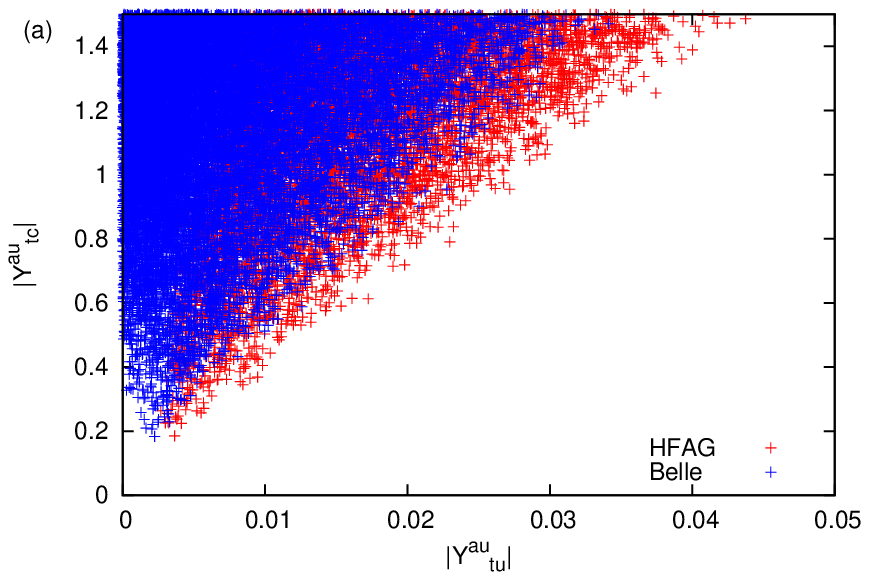,width=0.5\textwidth}}{\epsfig{figure=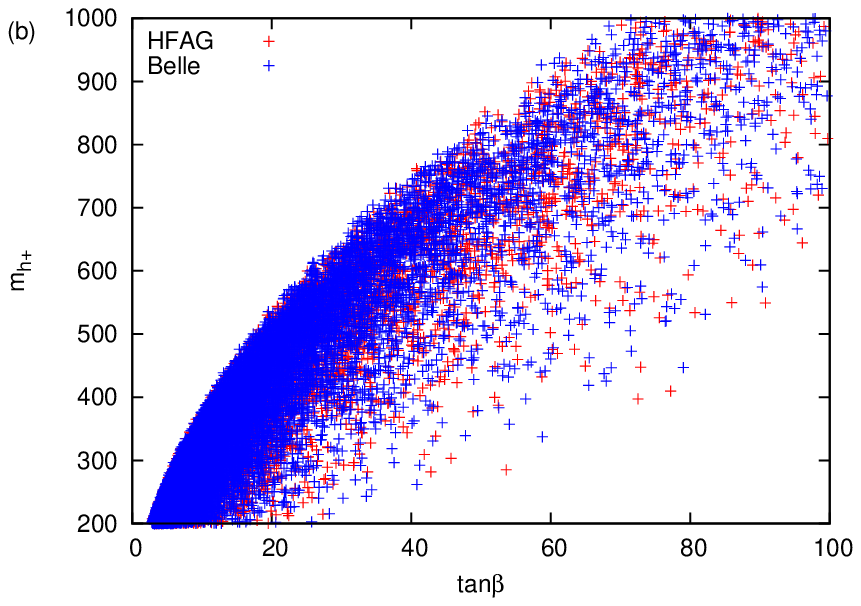,width=0.5\textwidth}}
\end{center}
\vspace{-0.5cm}
\caption{
Bounds on (a) $|Y^{au}_{tu}|$ and $|Y^{au}_{tc}|$ and
(b) $\tan \beta$ and $m_{h^+}$ in 2HDM. 
The points are consistent with $R(D^{(*)})$ within $1 \sigma$. 
The red points are consistent with the combined data
for $\textrm{BR}(B\to \tau \nu)$~\cite{Asner:2010qj}
while the blue points are in agreement with the new Belle data
for $\textrm{BR}(B\to \tau \nu)$~\cite{Adachi:2012mm}.
}
\label{fig:2HDM-Y}
\end{figure}

On the other hand, the BaBar data on $R(D)$ and $R(D^*)$~\cite{Crivellin:2012ye} 
prefers a large $C^{cb}_L/C^{cb}_{SM}$ (see eq.~(\ref{eq:CL}) and (\ref{eq:CR})). 
In figure~\ref{fig:2HDM-Y} (a) and (b), we show favored regions 
(a) for Yukawa couplings $|Y^{au}_{tu}|$ and $|Y^{au}_{tc}|$, 
and (b) $\tan\beta$ and $m_{h^+}$, which are consistent with $R(D)$ 
and $R(D^\ast)$ at BaBar within $1\sigma$, respectively. 
The red points are consistent with the combined data for
$\textrm{BR}(B\to \tau\nu)$ while the blue points agree with
the recent Belle data for $\textrm{BR}(B\to \tau\nu)$.
$|Y^{au}_{tu}|$ is restricted to be less than $0.05$
while $|Y^{au}_{tc}|$ is allowed to be $O(1)$. 
For the new Belle data $|Y^{au}_{tu}|$ is more constrained because
the data are more consistent with the SM prediction and leave little room for 
the charged Higgs contributions to $B \rightarrow \tau\nu$.
In order to account for the discrepancies in $R(D^{(*)})$,
the Yukawa coupling $|Y^{au}_{tc}|$ has to be sizable 
and its lower bound is about $0.2$. 
As we have discussed in figure~\ref{fig:2HDM-Btaunu}, $|Y^{au}_{tu}|$ of $O(1)$
might be consistent with $\textrm{BR}(B\to\tau \nu)$ experiments
if $R(D^{(*)})$ are not taken into account.   Basically $|Y^{au}_{tu}|$ is
constrained by $\textrm{BR}(B\to \tau\nu)$ while $|Y^{au}_{tc}|$ is by 
$R(D^{(*)})$. However they are related to each other through $\tan \beta$ and
$(g_R^u)$ (see eq.~(\ref{eq:Yuij_a})).   Hence, parameters which generate large 
$Y^{au}_{tu}$  are excluded by $R(D^{(*)})$ data at BaBar, and the large top FB
asymmetry at the Tevatron cannot be realized in our 2HDM. 
In figure~\ref{fig:2HDM-Y} (b), $\tan\beta \gtrsim 3$ is required in both
the combined and new Belle data. For large $\tan \beta$,
$m_{h^+}$ tends to be large. This is natural since $C_{L,R}^{cb}$
in (\ref{eq:CL}) and (\ref{eq:CR}) are proportional 
to $\tan\beta/m_{h^+}$ except for the last term in eq.~(\ref{eq:CL}).

Since only the RH up-type quarks are  charged non-universally under $U(1)'$, 
our models do not generate the dangerous  tree-level FCNC contributions to 
$B_d$-$\ov{B}_d$, $B_s$-$\ov{B}_s$, and $K^0$-$\ov{K^0}$ mixings. 
The $(u,c)$ elements of Yukawa couplings for neutral scalars and pseudoscalar,
which may generate a tree-level FCNC contributing to $D^0$-$\ov{D^0}$ mixing,
are small due to the suppression factor of the light-quark mass.
If a pseudoscalar has large $(t,u)$ and $(t,c)$ elements of Yukawa couplings,
they may enhance $D^0$-$\ov{D^0}$ mixing at the one-loop level. 
The loop of the pseudoscalar would induce the operators, 
$C_1 (\ov{u_R} \gamma^{\mu} c_R) (\ov{u_R} \gamma^{\mu} c_R)$ 
and $C_2 (\ov{u_R} c_L) (\ov{u_R} c_L)$, but the contribution to $C_1$ vanishes
if external momenta are set to be zero. 
Only $C_2$ is non-vanishing, but the contribution is 
suppressed by the factor, $m^2_c/m^2_{a}$. The upper bound 
on $C_2$ is discussed in ref.~\cite{Blum:2009sk, Gedalia:2009kh}: 
$|C_2| \lesssim 1.6 \times 10^{-7} $ TeV$^{-2}$.
Based on the general Higgs potential analysis in appendix~\ref{sec:potential}, 
the mass difference between the charged Higgs and the pseudoscalar boson is 
at most the week boson mass scale. Roughly speaking, $C_2$ could be estimated 
as $O((m^2_c/m^4_{a}) (Y^{au*}_{tu}Y^{au}_{tc})^2/( 16 \pi^2))$, 
which is much less than the upper bound on $C_2$.
Hence we can expect that the points in figure~\ref{fig:2HDM-Y} do not 
disturb the SM prediction in $D^0$-$\ov{D^0}$ mixing very much. 

As discussed in ref.~\cite{Ko-Top,Ko-Top2},  $Y^{au}_{tu} \sim O(1)$ is required 
to generate the large $A^t_{\rm FB}$ at the Tevatron and to evade the strong constraint from the same-sign
top-quark pair production at the LHC. However, figure~\ref{fig:2HDM-Y} tells 
that $Y^{au}_{tu}$ should be less than $5 \times O(10^{-2})$,  and 
$\tan \beta \gtrsim 3$ 
for $m_{h^+} > 200$ GeV 
\footnote{ The region $m_{h^+} < 200$ GeV is strongly constrained 
by $B \rightarrow X_s \gamma$ and search for the exotic top decay.}. 
There is also a strong  constraint from 
$\Delta \rho$.
According to eq.~(\ref{eq:Constraint-Rho}), the lower bound on $\tan \beta$ 
indicates $\Delta \rho_{\rm tree} \gtrsim 0.81 \times  
(g'^2/m_{\hat{Z}'}^2)(  m_{\Hat{Z}}^2/g^2_Z )$
in the limit, $m_{\hat{Z}}^2 \ll m_{\Hat{Z}'}^2$, in our 2HDM. 
This leads that the size of the $Z'$ interaction, 
$g'^2/m_{\hat{Z}'}^2$, should be by at least $O(10^{-3})$ times 
smaller than the size of the $Z$ interaction, $g_Z^2/m_{\hat{Z}}^2$, 
to achieve the $\Delta \rho$ within $1$ $\sigma$. 
This is too small to enhance the $A^t_{\rm FB}$ \cite{Ko-Top,Ko-Top2,Ko-LHC}.
This implies that we cannot consider
a sizable $Z'$ interaction while achieving the small $\Delta \rho$.

Therefore   in our 2HDM, it is difficult to find a favored region which is consistent 
with $R(D^{(*)})$ and $\textrm{BR}(B\to \tau\nu)$ as well as $A^t_{\rm FB}$ 
at the Tevatron.
If $R(D)$ and $R(D^*)$ become consistent with the SM prediction in the future,
a tiny $Y^{au}_{tc}$ is favored and the only constraint on $Y^{au}_{tu}$
will come from figure~\ref{fig:2HDM-Btaunu} (b).
According to figure~\ref{fig:2HDM-Y}, small $m_{h^+}/ \tan \beta$
is required by the large deviations of $R(D^{(*)})$ and large $Y^{au}_{tc}$,
so that a large $m_{h^+}/ \tan \beta$, where the new physics contribution is 
more suppressed,
could be chosen to realize a large $Y^{au}_{tu}$, if the deviations 
in $R(D^{*})$ become smaller. 

\section{3HDM \label{sec:3HDM}}

In order to enhance the $A^t_{\rm FB}$,  we need to have $O(1)$ $(t,u)$ elements of 
Yukawa couplings for both a CP-even scalar and a pseudoscalar~\cite{Ko-Top,Ko-Top2}. 
On the other hand, the $R(D^{(*)})$ data reported by the BaBar experiments required  
very tiny $Y^{au}_{tu}$   in our 2HDM, which cannot produce large  $A^t_{\rm FB}$. 
One simple solution to achieve both of the large $A^t_{\rm FB}$ and the BaBar 
discrepancies  is to consider 3HDM. In the 3HDM of chiral $U(1)'$ models, 
we have one more pair of charged Higgs and one more pseudoscalar, so that 
we can realize a scenario that one charged Higgs pair is constrained by $B$ decays, 
but the other decouples with the $B$ physics. In fact, 
we will find the parameter 
set that the Yukawa coupling of one pseudoscalar is $O(1)$.

\subsection{Yukawa couplings} 
In this section, we consider the 3HDM with the $( u_k ) = (-1,0,1)$ $U(1)'$ charge assignment of the RH up-type quarks, which was introduced in ref.~\cite{Ko-Top,Ko-Top2}.
In our 3HDM, there are $4$ CP-even neutral scalars $(h_1,h_2,h_3,h_4)$,  
$2$ CP-odd pseudoscalars $(a_1,a_2)$, and $2$ charged Higgs pair 
$(h_1^{\pm},h_2^{\pm})$, after the gauge symmetry breaking. 
The Goldstone modes which are eaten by $W$ and $Z$ bosons are 
linear to the VEVs, $( \langle H_1 \rangle,\,\! \langle H_2 \rangle,\,\! \langle H_3 \rangle)$,
and the orthogonal directions correspond to the mass eigenstates of pseudoscalars 
$(\xi^{a}_{1,2})$ and  charged Higgs $(\xi^{c}_{1,2})$.
Defining 
$( \langle H_1 \rangle,\,\! \langle H_2 \rangle,\,\! \langle H_3 \rangle)
= \frac{v}{\sqrt{2}}
(\sin \beta \cos \gamma,\,\!  \cos \beta,\,\! \sin \beta \sin \gamma  )$,  
we find that $\xi^{a,c}_{1,2}$ are given by 
\begin{eqnarray}
\xi^{a,c}_1&=&\cos \alpha_{a,c} 
\begin{pmatrix} -\cos \beta \cos \gamma \\  
\sin \beta \\ -\cos \beta \sin \gamma \end{pmatrix}
+\sin \alpha_{a,c} 
\begin{pmatrix} \sin \gamma \\ 0 \\ - \cos \gamma  \end{pmatrix}, \\
\xi^{a,c}_2&=&-\sin \alpha_{a,c} 
\begin{pmatrix} -\cos \beta \cos \gamma \\  \sin \beta \\ 
-\cos \beta \sin \gamma \end{pmatrix}
+\cos \alpha_{a,c} 
\begin{pmatrix} \sin \gamma \\ 0 \\ - \cos \gamma  \end{pmatrix}. 
\end{eqnarray} 
The mixing parameters, $\alpha_{a,c}$, relate to the terms, 
$M_{ij}$ and $\widetilde{\lambda}_{ij}$, in the Higgs potential, 
as described in appendix~\ref{sec:potential}.   
The Yukawa couplings of pseudoscalars are given by
\begin{eqnarray}\label{eq:3HDM-pseudo}
Y^{au(1)}_{ij}&=&-\Hat{Y}^{u(1)}_{ij} \cos \alpha_a  
+ \Hat{Y}^{u(2)}_{ij}  \sin \alpha_a,  \\
Y^{au(2)}_{ij}&=&\Hat{Y}^{u(2)}_{ij}  \cos \alpha_a  
+\Hat{Y}^{u(1)}_{ij} \sin \alpha_a,   
\end{eqnarray}
where $\Hat{Y}^{u(1)}_{ij}$ and $\Hat{Y}^{u(2)}_{ij}$ are defined as
\begin{eqnarray}\label{eq:3HDM-pseudo2}%
\Hat{Y}^{u(1)}_{ij}&=&\frac{m_i^u}{v} \left \{ \frac{1}{\tan \beta} \delta_{ij} 
- \frac{2}{\sin 2 \beta}  (R_{i2} R^*_{j2}) \right \},  \\
\Hat{Y}^{u(2)}_{ij}&=&\frac{m_i^u}{v} 
\left \{ \frac{\tan \gamma}{\sin \beta} \delta_{ij}
- \frac{\tan \gamma}{\sin \beta}  (R_{i2} R^*_{j2})
- \frac{2}{\sin 2 \gamma}  (R_{i3} R^*_{j3})  \right \}. 
\label{eq:pseudo2-3HDM}%
\end{eqnarray}
The down-type quark sector and lepton sector Yukawa couplings are
\begin{eqnarray}
Y^{ad(1)}_{ij}&=&  \delta_{ij} \frac{m_i^d}{v} \tan \beta \cos \alpha_a,  \\
Y^{ad(2)}_{ij}&=& - \delta_{ij} \frac{m_i^d}{v} \tan \beta \sin \alpha_a, \\
Y^{al(1)}_{ij}&=&  \delta_{ij} \frac{m_i^l}{v} \tan \beta \cos \alpha_a,  \\
Y^{al(2)}_{ij}&=& - \delta_{ij} \frac{m_i^l}{v} \tan \beta \sin \alpha_a.   
\end{eqnarray}

The magnitude of off-diagonal elements of $\Hat{Y}^{u(1)}_{ij}$ 
is  the same as the corresponding Yukawa coupling in our 2HDM
except for replacement of $(R_{i2} R^*_{j2})$ by $(g^u_R)_{ij}$.
The Yukawa couplings of the charged Higgs are obtained from the relation
(\ref{eq:relation}) with replacement of $\alpha_a$ by $\alpha_c$ 
in $Y^{au(1,2)}_{ij}$, $Y^{ad(1,2)}_{ij}$, and $Y^{ae(1,2)}_{ij}$.
 
\subsection{ $R(D^{(*)})$ and $\textrm{BR}(B \to\tau \nu)$ in 3HDM}
Now let us consider $R(D^{(*)})$ and $\textrm{BR}(B \to\tau \nu)$ 
in our 3HDM. There are two charged Higgs pairs that contribute to $C^{qb}_L$ and 
$C^{qb}_R$, which are estimated at the charged Higgs mass scale as follows:
\begin{eqnarray}
\frac{C^{qb}_L}{C^{qb}_\textrm{SM}}&=& v m_{\tau} \tan \beta 
\left ( \frac{V^*_{kb}}{V_{qb}} \right ) \left \{ - \Hat{Y}^{u(1)}_{kq}  
\left ( \frac{ \cos^2 \alpha_c}{m^2_{h_1^+}} 
+ \frac{ \sin^2 \alpha_c}{m^2_{h_2^+}} \right ) 
+ \Hat{Y}^{u(2)}_{kq}   \frac{\sin 2 \alpha_c}{2} \left ( \frac{1}{m^2_{h_1^+}}
- \frac{1}{m^2_{h_2^+}} \right ) \right \}, \nonumber \\ 
&& \\
\frac{C^{qb}_R}{C^{qb}_\textrm{SM}}&=&- m_b m_{\tau} \tan^2 \beta 
\left ( \frac{ \cos^2 \alpha_c }{m^2_{h_1^+}} 
+ \frac{  \sin^2 \alpha_c }{m^2_{h_2^+}   } \right ). 
\end{eqnarray}      
$R(D^{(\ast)})$ and $\textrm{BR}(B\to \tau\nu)$ depend on several parameters.
In order to find  the regions favored by the BaBar and Belle data,
we vary each parameter in the following range:
$1 \le \tan \beta \le 100$,
$200~\textrm{GeV} \le m_{h_1^+} \le 1~\textrm{TeV}$,
$200~\textrm{GeV} \le m_{h_2^+} \le 400~\textrm{GeV}$,
$0 \le \alpha_{a,c} \le 2\pi$,
and
$0\le |R_{i2}R_{j2}^\ast|, |R_{i3}R_{j3}^\ast| \le 1$, respectively,
in addition to constraints on the Yukawa couplings,
$|Y_{tu(tc)}^{u(1,2)}|\le 1.5$. 
In the numerical analyses, we take $\tan \gamma=1$ in order to realize a small tree-level 
contribution to the $\rho$-parameter \footnote{Note that eq.~(\ref{eq:Constraint-Rho}) 
and $\{ h_i (\langle H_i \rangle \sqrt{2} /v )^2 \} \propto (1- \tan^2 \gamma)$ in the 3HDM.}.
After imposing the experimental constraints from  $R(D^{(\ast)})$
at BaBar~\cite{Lees:2012xj}, $\textrm{BR}(B\to \tau \nu)$~\cite{Asner:2010qj},
and the $D^0$-$\ov{D^0}$ mixing, we could find parameter regions consistent with 
all the experimental constraints. For the bound on $\textrm{BR}(B\to \tau \nu)$,
we use the HFAG value, because the result does not change much
even if we adopt the new data at Belle.

Let us discuss a few specific cases for simplicity. First, we consider the case where 
$\cos \alpha_c = 1$. In this case, $C^{qb}_{L,R}/C^{qb}_{SM}$ do not depend 
on $m_{h_2^+}$, so that $h_2^+$
 decouples from 
$B\to D^{(\ast)}\tau \nu$ and $B\to \tau \nu$, 
but we note that the contribution of the pseudoscalar exchanges to 
the $D^0$-$\ov{D^0}$ mixing diagram also constrains the model parameters.
Then, we can apply the same discussion as in the 2HDM
by replacing $(R_{i2} R^*_{j2})$ with $(g^u_R)_{ij}$.
In this case, $|\Hat{Y}^{u(1)}_{tu}(R_{t2} R^*_{u2})|
=|\Hat{Y}^{au}_{tu}(R_{t2} R^*_{u2})|$ is satisfied, so that $|\Hat{Y}^{u(1)}_{tu}|$ 
should be smaller than $0.05$ according to the discussion in our 2HDM.
On the other hand, $\Hat{Y}^{u(2)}_{tu}$ depends not only on $(R_{t2} R^*_{u3})$ 
but also on $(R_{t3} R^*_{u3})$ (see eq.~(\ref{eq:pseudo2-3HDM})).
Since the combination $(R_{i3} R^*_{j3})$ certainly enhances $\Hat{Y}^{u(2)}_{tu}$, 
the exchange of the heavier pseudoscalar boson $a_2$ 
could realize the large 
$A^t_{\rm FB}$ when $\cos \alpha_a $  is also $1$.\footnote{In appendix~\ref{sec:potential}, the condition for $\cos \alpha_c=\cos \alpha_a =1$ is discussed.}

\begin{figure}[!t]
\begin{center}
{\epsfig{figure=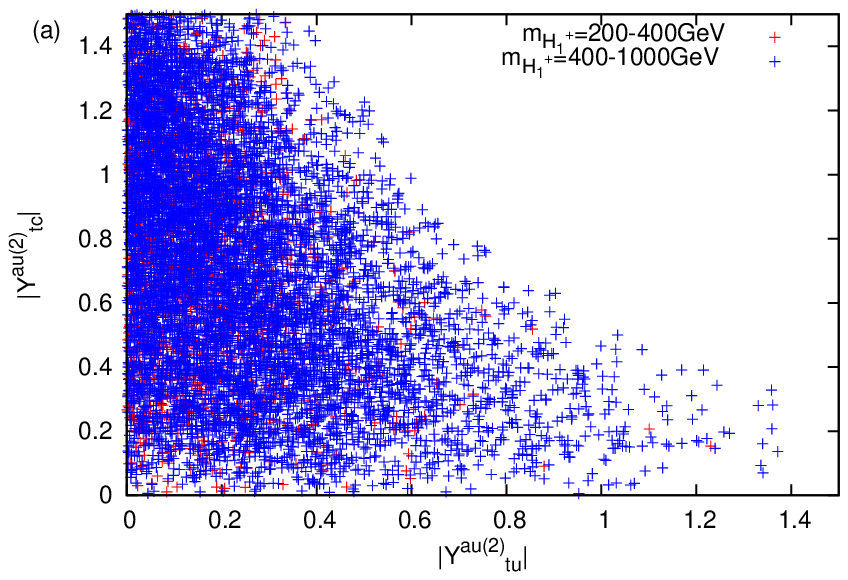,width=0.5\textwidth}}%
{\epsfig{figure=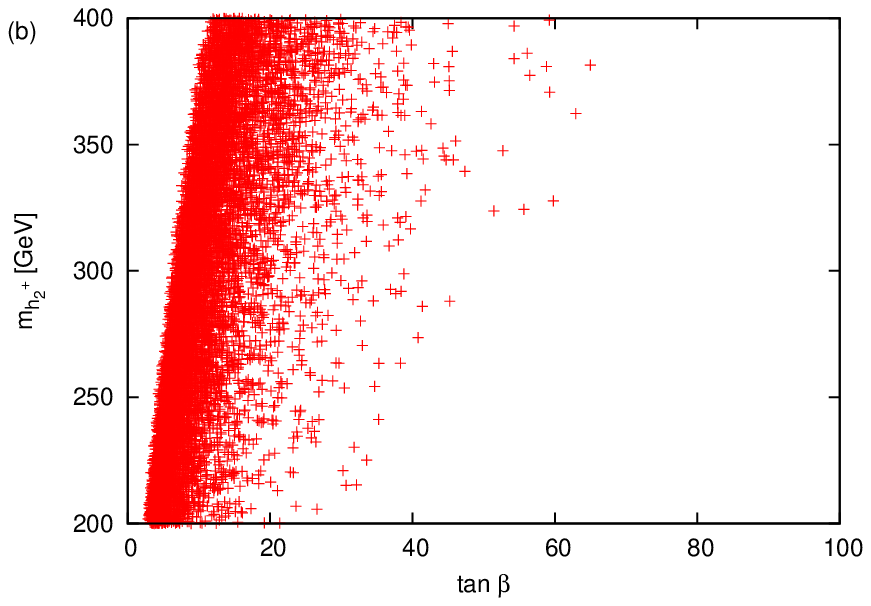,width=0.5\textwidth}}  
\end{center}
\vspace{-0.5cm}
\caption{
(a) $|Y^{au(2)}_{tc}|$ vs. $|Y^{au(2)}_{tu}|$ and
(b) $\tan \beta$ vs. $m_{h_2^+}$ for $\sin\alpha_c=1$ in 3HDM.
The red (blue) points in (a) are allowed points 
for $200~(400)~\textrm{GeV} \le m_{h_1^+} \le 400~(1000)~\textrm{GeV}$.
}
\label{fig:3HDM2}
\end{figure}

Secondly, we consider the case where $\sin\alpha_c=1$. 
In this case, $C_{R,L}^{qb}/C_\textrm{SM}^{qb}$ is independent of $h_1^+$. 
Then $h_1^+$ decouples from $B\to D^{(*)}\tau \nu$ and $B\to \tau\nu$.
In figure~\ref{fig:3HDM2}, we depict $|Y^{au(2)}_{tc}|$ vs. $|Y^{au(2)}_{tu}|$
and $\tan \beta$ vs. $m_{h_2^+}$ for $\sin\alpha_c=1$ with the constraints from the $D^0$-$\ov{D^0}$ mixing. 
In figure~\ref{fig:3HDM2} (a), the red points are allowed points 
for $200~\textrm{GeV} \le m_{h_1^+} \le 400~\textrm{GeV}$ while
the blue ones are for $400 ~\textrm{GeV} \le m_{h_1^+} \le 1 ~\textrm{TeV}$,
respectively. 
As we see in figure~\ref{fig:3HDM2} (a), we can find 
the allowed points with small $|Y^{au(2)}_{tc}|$ and 
$|Y^{au(2)}_{tu}|$ of $O(1)$, which are in agreement with 
$R(D^{(*)})$ and $\textrm{BR}(B\to \tau\nu)$.
As we have already discussed, $|Y^{au(2)}_{tu}|$ of $O(1)$ might realize the 
large $A_\textrm{FB}^t$ at the Tevatron. 
In this scenario, $h_2^+$ contributes to the $B$ decays and $a_2$ 
could enhance the $A_\textrm{FB}^t$.
As we see in figure~\ref{fig:3HDM2} (b), 
$\tan\beta$ tends to increase slightly as $m_{h_2^+}$ becomes large.
Hence, small $\tan\beta$ is also favored to increase the enhancement,
because the mass difference between $h_2^+$ and $a_2$ would be an order 
of electroweak scale.  

\begin{figure}[!t]
\begin{center}
{\epsfig{figure=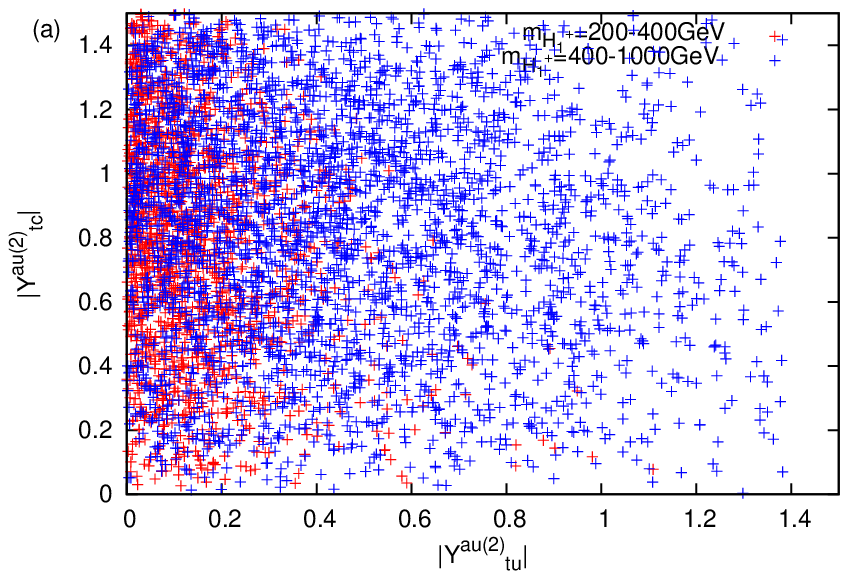,width=0.5\textwidth}}%
{\epsfig{figure=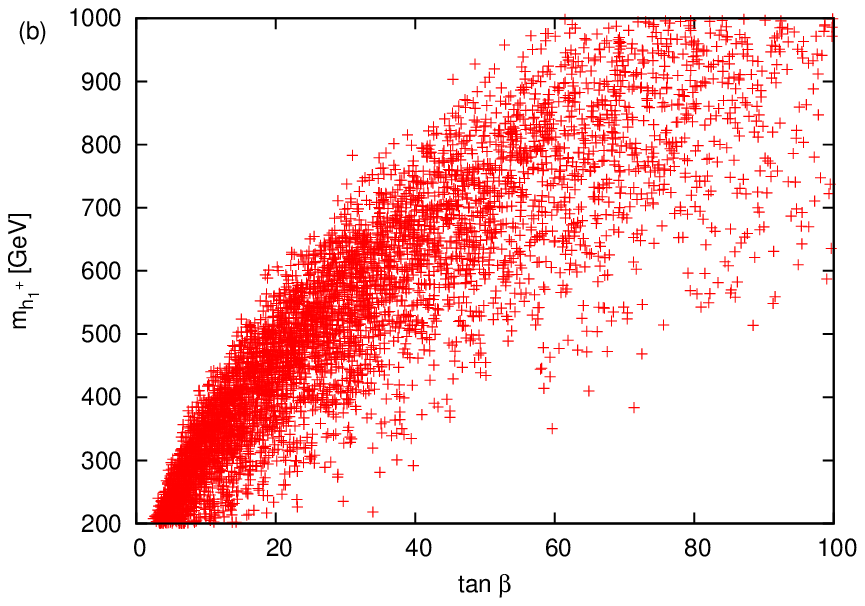,width=0.5\textwidth}}  
\end{center}
\vspace{-0.5cm}
\caption{
(a) $|Y^{au(2)}_{tc}|$ vs. $|Y^{au(2)}_{tu}|$ and 
(b) $\tan \beta$ vs. $m_{h_1^+}$ for $m_{h_1^+}=m_{h_2^+}$ in 3HDM.
The red (blue) points in (a) are allowed points 
for $200~(400)~\textrm{GeV} \le m_{h_1^+} \le 400~(1000)~\textrm{GeV}$.
}
\label{fig:3HDM3}
\end{figure}

Finally, we consider the degenerate case of the charged Higgs pairs with
$m_{h_1^+}=m_{h_2^+}$.
In this case, $C_L^{qb}/C_\textrm{SM}^{qb}$ and $C_R^{qb}/C_\textrm{SM}^{qb}$ 
are independent of $\alpha_c$ and $\hat{Y}_{kq}^{u(2)}$.
We show the scattered plots for $|Y^{au(2)}_{tu}|$ vs. $|Y^{au(2)}_{tc}|$
in figure~\ref{fig:3HDM3} (a) and for $\tan\beta$ vs. $m_{h_1^+}$
in figure~\ref{fig:3HDM3} (b), respectively.
We find that the tendency between $\tan\beta$ and $m_{h_1^+}$ looks similar to
that in 2HDM or in the $\cos\alpha_c=1$ case in 3HDM.
As discussed in ref. \cite{Ko-Top,Ko-Top2, Ko-LHC,Ko-LHC2}, the small mass of the pseudoscalar of around $300$ GeV,
is favored to enhance  $A_\textrm{FB}^t$. 
As we see in figure~\ref{fig:3HDM3} (a), 
one can find a few red points with $O(1)$ $|Y^{au(2)}_{tu}|$ and light $m_{h_1^+}$
which could realize
 the large $A_\textrm{FB}^t$ at the Tevatron.
If  $m_{h_1^+}$ is heavier than $400$ GeV, very large $\widetilde{\lambda}_{ij}$
in the Higgs potential is required as discussed in appendix~\ref{sec:potential}. 

In conclusion, we can find the parameter regions which could explain
the deviation of $R(D^{(*)})$ in the BaBar experiment and be consistent with
$\textrm{BR}(B\to \tau\nu)$ at the $B$ factories in 3HDM. We further investigate
three interesting cases, (i) $\cos\alpha_c=1$, (ii) $\sin\alpha_c=1$,
and (iii) $m_{h_1^+}=m_{h_2^+}$. In all the three cases, we could 
find the allowed regions with large $|Y^{au(2)}_{tu}|$, which in turn  could 
generate large top forward-backward asymmetry at the Tevatron.

\section{The Other Constraint on $m_{h^+}$
\label{sec:constraints}}

It is well known that $B \to X_s \gamma$ can give a stringent bound on the 
charged Higgs mass depending on the details of models.  
In the type-II 2HDM, the lower bound of $m_{h^+}$
is about $300$ GeV
at next-to-next-to-leading order~\cite{Misiak1,Misiak2,Becher:2006pu,Misiak3}. 
In our multi-Higgs doublet models with flavor-dependent $U(1)'$ gauge interactions, 
the Yukawa couplings of the charged Higgs  have extra parameters.
The $B \to X_s \gamma$ decay occurs through the loop diagram involving the top 
quark  and the charged Higgs boson,   where the relevant element of Yukawa 
couplings is the $(b,t)$ element.   According to eq.~(\ref{eq:relation}), we can expect 
that the $(b,t)$ element of the charged  Higgs is governed by the $(t,t)$ elements of 
the pseudoscalar bosons.   In principle, the $(t,t)$ elements of the pseudoscalar 
bosons have other mixing parameters,  such as $(g_R^u)_{tt}$ in our 2HDM,  which,
however, are not directly constrained by the semileptonic and leptonic $B$ decays.   
There is a theoretical relation, $|(g_R^u)_{tq}|^2=(g_R^u)_{qq}(g_R^u)_{tt} (q=u,c)$, 
in our 2HDM,  so that $O(1)$ $(g_R^u)_{tc}$ for $R(D^{(*)})$ requires $O(1)$ 
$(g_R^u)_{tt}$.  

The bound on $B \to X_s \gamma$ at leading order (LO)
up to $O((100~{\rm GeV}/m_{h^+})^2)$ is given by
\begin{equation}
\label{eq:bsgamma}
-0.20 \lesssim \left \{ - \left( 46.26+46.83 \ln \left ( \frac{100~{\rm GeV}}{m_{h^+}} \right )  
\right )  Y^{au}_{tt} \tan \beta + 9.00 (Y^{au}_{tt})^2  \right \} 
\left( \frac{100~{\rm GeV}}{m_{h^+}} \right)^2  \lesssim 0.79,
\end{equation}
where two relations $Y_{bt}^{-u}=\sqrt{2} V_{tb} Y^{au}_{tt}$ and 
$Y_{tb}^{+d}=\sqrt{2} V_{tb} Y^{ad}_{bb} =  m_b \tan \beta /v$ are used
~\cite{Borzumati:1998tg}.
If we assume $\tan \beta=1$ and $m_{h^+}=300$ GeV, then we obtain 
a constraint  $-0.077 \lesssim Y^{au}_{tt} \lesssim 0.262$. 
Therefore we can expect that $(g_R^u)_{tt}$ can be  $O(1)$ without conflict with 
the $B\rightarrow X_s \gamma$ constraint.

\section{Summary  \label{sec:summary}}

In this paper, we investigated the constraints from the semileptonic and leptonic 
$B$ decays on our 2HDM and 3HDM, which were proposed in ref.~\cite{Ko-Top,Ko-Top2,Ko-LHC,Ko-LHC2} in order to accommodate the top forward-backward asymmetry 
$(A^t_{\rm FB})$  observed at the Tevatron.   In ref.~\cite{Ko-Top,Ko-Top2,Ko-LHC,Ko-LHC2}, 
the $U(1)'$-charged extra Higgs doublets were introduced in order to generate the 
realistic Yukawa couplings for the up-type quarks (especially the top quark mass)
~\cite{Ko-Top,Ko-Top2}. In the previous study, not only CP-even scalar bosons but also 
pseudoscalar bosons are required to have $O(1)$ $(t,u)$-element Yukawa couplings 
in order to achieve the large $A^t_{\rm FB}$ and  to evade the strong bound from the same-sign top-quark pair production signal at the LHC~\cite{Ko-Top,Ko-Top2,Ko-LHC,Ko-LHC2}.
On the other hand, such a large  $(t,u)$ element of the pseudoscalar Yukawa 
coupling allows a large $(b,u)$ element to appear  in the charged 
Higgs Yukawa couplings. This implies that our model might predict large 
deviations from the SM predictions in $B$ physics.
For example, the $(t,u)$ element of the pseudoscalar boson is 
constrained indirectly by the $B \to \tau \nu$ decay.

Recently, the BaBar collaboration reported the interesting results for 
$R(D^{(*)})$ in the semileptonic $B$ decays, $B \to D^{(*)} \tau \nu$. 
The combined results deviate from the SM predictions by $3.4\sigma$ 
and require large FCNCs beyond MFV if the signal comes from charged 
Higgs exchanges.  Our $U(1)'$-flavored multi-Higgs doublet models naturally 
realize a large $(b,c)$ element of the charged Higgs. 
Therefore in this paper, we investigated if our models explain both the BaBar 
discrepancies and $A^t_{\rm FB}$ at the Tevatron.

In our 2HDM, only one Higgs doublet is charged under $U(1)'$, so that small 
$\tan \beta$ is required to avoid the constraint on the tree-level $\rho$ parameter. 
The $R(D^{(*)})$ discrepancies at BaBar
require large new physics effects, so that small
$m_{h^+}/\tan \beta$ is required as we see in figure~\ref{fig:2HDM-Y} (b).
However, a small $m_{h^+}/\tan \beta$ requires a very small $|Y^{au}_{tu}|/\tan \beta$
according to the bound from $B \to \tau \nu$ as shown in figure~\ref{fig:2HDM-Btaunu} (b), 
so that we could not find the points  in figure~\ref{fig:2HDM-Y} (a) with large 
$Y^{au}_{tu}$ which is needed for large $A^t_{\rm FB}$.
If $R(D^{(*)})$ converge to the SM prediction in the future, we would not need consider such large new physics effect and  large $Y^{au}_{tc}$, so that we 
would be able to choose the points with large $Y^{au}_{tu}$ and small 
$\tan \beta$ in figure~\ref{fig:2HDM-Btaunu} (b).

In our 3HDM, we have more freedom: one more charged Higgs pair 
and one more pseudoscalar boson. 
It is not difficult to find the allowed points for $R(D^{(*)})$
and $\textrm{BR}(B\to \tau \nu)$ in the general case.
In the limit, $\cos \alpha_c \rightarrow 1$, 
one of the charged Higgs bosons does 
not contribute to the (semi)leptonic $B$ decay,
so that we can describe the scenario that 
one of the charged Higgs bosons
explains the BaBar results and the other becomes independent 
of the $B$ physics. 
The explicit relation like eq.~(\ref{eq:relation}) is not respected 
in this 3HDM because of the extra freedom, $\alpha_{a,c}$. 
We also investigated the scalar potential in appendix~\ref{sec:potential}  
and derived the condition for $\cos \alpha_{a,c} = 1$, where 
both of $A^t_{\rm FB}$ at the Tevatron and $R(D^{(*)})$ at BaBar could be achieved. 
In figures~\ref{fig:3HDM2} and \ref{fig:3HDM3}, we also plotted the allowed 
region, setting $\tan \gamma=1$ 
which corresponds to $\Delta \rho_{\rm tree}=0$, and discussed 
the two cases: $\sin \alpha_c=1$ and $m_{h_1^+}=m_{h_2^+}$. 
We can find the points with $O(1)$ (t,u) Yukawa coupling and $300-400$ GeV 
charged Higgs mass, which are consistent 
with $R(D^{(*)})$ and $D^0$-$\ov{D^0}$ mixing. $\tan \beta \lesssim 40$ 
should be satisfied in $\sin \alpha_c=1$ case, and the pair, $(h_2^+,a_2)$, could
achieve the BaBar discrepancies and $A^t_{\rm FB}$. 
In the degenerate limit, 
$m_{h_1^+}=m_{h_2^+}$, the $\alpha_c$ dependence disappears 
in the $B$ decays and we find the tendency between $\tan \beta$ and $m_{h_1^+}$
as shown in figure~\ref{fig:3HDM3}.
However, the new contributions from the extra scalars are also constrained 
by the  $D^0$-$\ov{D^0}$ mixing,
so that only a few points with a large $(t,u)$ Yukawa coupling are allowed if $m_{h_1^+}$ is lighter than $400$ GeV.

We also commented on the other bound on the charged Higgs boson. 
One of the most important bounds on the charged Higgs boson comes from
$B \to X_s \gamma$. 
In our models, the $(t,t)$ element of Yukawa coupling contributes 
to the process at LO and we derived the bound in the limit 
that the charged Higgs boson is heavy. 
As discussed in ref.~\cite{Misiak1,Misiak2,Becher:2006pu,Misiak3},
the lower bound on the charged Higgs mass in the type-II 2HDM is around $300$ GeV, 
but the mass below the bound is possible in our models, depending on the $(t,t)$ element.

Before closing this paper, let us emphasize once more the importance to study 
phenomenology in the framework of a well defined consistent renormalizable 
lagrangian.  The original $Z'$ model for the top forward-backward asymmetry has been excluded 
several times by the upper bound on the same-sign top-quark pair production cross section. 
However the model with only $Z'$ is not well defined since it is not renormalizable
and not realistic because the up-type quarks including top quark are massless. 
It is mandatory to extend the Higgs sector, by introducing new Higgs doublets with
nonzero $U(1)'$ charges. Making such an extension actually affects the top phenomenology a lot.  Basically all the top-related observables are affected by 
extra Higgs doublets, both neutral and charged Higgs bosons. Also the $B$ meson
and $D$ meson sectors are modified too.  Maybe it is timely to remind ourselves that
the SM with three quarks $u,d,s$ produces too large FCNCs in the kaon sector, 
and is immediately excluded. This disaster can be cured only by introducing the 
4th quark, the charm quark, which also solves the problem of gauge anomaly. 
The experience with the charm quark already teaches us that it is important to 
work  in a minimal consistent (anomaly free) and renormalizable model 
in order to do a meaning phenomenology.

\appendix

\section{Potential Analysis}
\label{sec:potential}
The general potential in multi-Higgs-doublet models with $U(1)'$ is given by
\begin{eqnarray}
V&=& m^2_i H^{\dagger}_i H_i - \left ( m^2_{ij} (\Phi) H^{\dagger}_i H_j 
+h.c.  \right )  \nonumber \\
&&+ \frac{\lambda_i}{2} (H^{\dagger}_i H_i)^2 
+ \lambda_{ij} (H^{\dagger}_i H_i)(H^{\dagger}_j H_j)  
+ \widetilde{\lambda}_{ij} (H^{\dagger}_i H_j)(H^{\dagger}_j H_i),
\end{eqnarray}
where $m^2_{ii} (\Phi)= \lambda_{ii}=\widetilde{\lambda}_{ii}=0$, 
$\lambda_{ij}=\lambda_{ji}$ and 
$\widetilde{\lambda}_{ij}=\widetilde{\lambda}_{ji}$ are satisfied. 
$m^2_{ij} (\Phi)$ is the function of $\Phi$, which is the complex scalar to 
break $U(1)'$ and the function is fixed by the charge assignment of the fields.
Each $H_i$ includes neutral, pseudoscalars and charged Higgs,
\begin{equation}
H_i = \begin{pmatrix} \phi^+_i  \\ 
\frac{v_i}{\sqrt{2}} + \frac{1}{\sqrt{2}} (h_i + i \chi_i ) \end{pmatrix}.
\end{equation}
$v_i$ satisfies the stationary condition,
\begin{equation}
0=m^2_i v_i - (m^2_{ij}+ m^{2*}_{ji})v_j + \lambda_i \frac{v^3_i}{2} 
+ \lambda_{ij}\frac{v_iv_j^2}{2} +  \widetilde{\lambda}_{ij} \frac{v_iv_j^2}{2}.
\end{equation}
Assuming $\Phi$ gets nonzero VEV, the mass matrices of pseudoscalars 
and charged Higgs are
\begin{eqnarray}
(M^2_{a})_{ij}&=&-M^2_{ij} + \frac{v_k}{v_i} M^2_{ik} \delta_{ij},   \\
(M^2_{h^+})_{ij}&=& (M^2_{a})_{ij} + \widetilde{\lambda}_{ij} 
\frac{v_iv_j}{2} - \widetilde{\lambda}_{ik} \frac{v^2_k}{2}\delta_{ij},
\end{eqnarray}
where $M^2_{ij}=m^2_{ij}+ m^{2*}_{ji}$ and we assume that $M^2_{ij}$ is real.

In 2HDM, we can find one simple relation between $m^2_{h^{+}}$ and $m^2_a$,
\begin{equation}
m^2_{h^{+}}=m^2_a -  \widetilde{\lambda}_{12} \frac{ v^2}{2}.
\end{equation}
This means that the mass difference is at most the electroweak scale.

In 3HDM, the directions of massive modes 
in pseudoscalar and charged Higgs sectors 
can not be fixed,
as we discuss in section~\ref{sec:3HDM}. The condition for $\cos \alpha_{c,a}=1$ is 
obtained from the following calculation,
\begin{eqnarray}
M^2_{a} \begin{pmatrix} \sin \gamma \\ 0 \\ - \cos \gamma \end{pmatrix} &=& 
\begin{pmatrix} 
M^2_{12} \frac{ \tan \gamma}{\tan \beta} + M^2_{13}  \frac{1 }{\cos \gamma} \\ 
-M^2_{12}  \sin \gamma +  M^2_{23} \cos \gamma  \\   
-M^2_{13} \frac{1}{ \sin \gamma} - M^2_{23}   \frac{1 }{\tan \beta \tan \gamma}
\end{pmatrix},   \\
M^2_{h^+} \begin{pmatrix} \sin \gamma \\ 0 \\ - \cos \gamma \end{pmatrix} &=&
M^2_{a} \begin{pmatrix} \sin \gamma \\ 0 \\ - \cos \gamma \end{pmatrix} 
- \frac{v^2}{2}  
\begin{pmatrix} \sin \gamma ( \widetilde{\lambda}_{12} \cos^2 \beta 
+ \widetilde{\lambda}_{13} \sin^2 \beta ) \\ 
( \widetilde{\lambda}_{23}
- \widetilde{\lambda}_{12}) \cos \beta \sin \beta \cos \gamma \sin \gamma \\ 
- \cos \gamma ( \widetilde{\lambda}_{23} \cos^2 \beta 
+ \widetilde{\lambda}_{13} \sin^2 \beta )  \end{pmatrix}.
\end{eqnarray}
That is, $M^2_{12}  \tan \gamma =  M^2_{23} \cos \gamma $ 
and $ \widetilde{\lambda}_{23}=\widetilde{\lambda}_{12}$ are required.

\acknowledgments
We thank Korea Institute for Advanced Study for providing computing resources (KIAS
Center for Advanced Computation Abacus System) for this work.
This work is supported in part by Basic Science Research Program through the
National Research Foundation of Korea (NRF) funded by the Ministry of Education Science
and Technology 2011-0022996 (CY), by NRF Research Grant 2012R1A2A1A01006053 (PK
and CY), and by SRC program of NRF funded by MEST (20120001176)
through Korea Neutrino Research Center at Seoul National University (PK).
The work of YO is financially supported by the ERC Advanced Grant project “FLAVOUR” (267104).


\end{document}